\documentclass[twocolumn,showpacs,amsmath,amssymb,prb,superscriptaddress]{revtex4-1}
\usepackage{graphicx}
\usepackage{bm}
\usepackage{dcolumn}
\usepackage{color}
\usepackage{epsfig}
\usepackage{epstopdf}

\begin{document}
\title{La$_2$O$_3$Fe$_2$Se$_2$, a Mott insulator on the brink of
  orbital-selective metalization}

\author{Gianluca Giovannetti}
\affiliation{CNR-IOM-Democritos National Simulation Centre, UoS
  Trieste-SISSA, Via Bonomea 265, I-34136, Trieste,
Italy}
\affiliation{International School
for Advanced Studies (SISSA), Via Bonomea 265, I-34136, Trieste, Italy}
\author{Luca de'~Medici}
\affiliation{ European Synchrotron Radiation Facility, BP 220, F-38043 Grenoble Cedex 9, France }
\author{Markus Aichhorn}
\affiliation{Institute of Theoretical and Computational Physics, TU
  Graz, Petersgasse 16, Graz, Austria}
\author{Massimo Capone}
\affiliation{International School
for Advanced Studies (SISSA), Via Bonomea 265, I-34136, Trieste,
Italy}
\affiliation{CNR-IOM-Democritos National Simulation Centre, UoS
  Trieste-SISSA, Via Bonomea 265, I-34136, Trieste,
Italy}

\begin{abstract}
We show that the insulating character of the iron-selenide
%Using a combination of density functional theory, slave-spins and dynamical
%mean-field theories we investigate the electronic structure of the recently
%synthesized insulator 
 La$_2$O$_3$Fe$_2$Se$_2$ can be explained in terms of Mott
localization in sharp contrast with the metallic behavior of FeSe and
other parent parent compounds of iron superconductors. 
We demonstrate that the key ingredient that makes
La$_2$O$_3$Fe$_2$Se$_2$ a Mott insulator, rather than a correlated
metal dominated by the Hund's coupling is the enhanced crystal-field splitting,
accompanied by a smaller orbital-resolved kinetic energy.
The strong deviation from orbital degeneracy introduced by the
crystal-field splitting also pushes this materials close to an
orbital-selective Mott transition. We predict that either doping or
uniaxial external pressure can drive the material into an
orbital-selective Mott state, where only one or few orbitals are
metallized while the others remain insulating.

\end{abstract}

\maketitle

\section{Introduction}
The link between high-temperature superconductivity and strong
electron-electron correlations has been forged and strengthened by decades of
investigation in the copper-based superconductors (cuprates).
In this light, the debate about the strength and the role of electron
correlations in iron-based superconductors (FeSC) maintains a crucial importance. 
The overall phenomenology of these materials does not provide a
self-evident answer. In these materials superconductivity appears
doping a metallic spin-density-wave parent compound, rather than the
Mott insulator of the cuprates, and, while the metallic state is
highly incoherent, the standard fingerprints of
strong correlations, like the Hubbard bands, are not universally observed.

Furthermore non-perturbative studies of the effect of the interactions, mainly based on Dynamical Mean-Field Theory (DMFT)
and related methods, have highlighted a novel behavior, in which the Hund's coupling $J_h$ to play a
major role in determining the degree of
correlations\cite{haule2009,aichhorn_theoretical_2010}. The electrons in the d-orbitals (the parent compounds have a
nominal filling of six electrons for each iron atom) are strongly
correlated, as measured by the bad-metallic behavior with small coherence
scales shown in many experiments, but the Hubbard repulsion $U$ is
substantially smaller than the critical value for the Mott transition\cite{demedici2011,lanata_orbital_2013}. This regime is often labeled as a ``Hund's
metal'' and displays characteristic properties\cite{gabi_power_laws,hansmann}
including a remarkable tendency to ``orbital selectivity'', i.e., to a
neat differentiation in the degree of correlation of the different
orbitals, leading to the simultaneous presence of weakly and strongly
correlated electrons\cite{demedici_genesis,KouLiWeng2009,*HacklVojta2009,*YinLeeKu2010,*PPhillips2010,*yin_kinetic_2011,*YuSi2011,*Bascones2012,*Valenti2012,YuSi_LDA-SlaveSpins_LaFeAsO,lanata_orbital_2013,demedici_selective2014}.
In Ref. \cite{demedici_selective2014} it has been shown that the
Hund's coupling decouples the orbitals quenching the inter-orbital
fluctuations and that this leads to a picture of five doped
single-band Mott insulators. 
Consequently the degree of correlation is
controlled by the distance of each individual orbital from the
half-filled d$^5$ configuration. The relevance of the $d^5$ Mott phase
in the phase diagram of models for iron superconductors has been
observed also in  \onlinecite{Liebsch2010,*Ikeda2010,*Imada2012}.

While no iron-based material can be doped with one hole per iron site, 
isostructural materials where iron is
replaced with manganese are characterized by the $d^5$ half-filled configuration
and they are indeed antiferromagnetic Mott
insulators.\cite{singh_bamn2as2_2009,yanagi2009,pandey_bamn2as2doped_2012}.
% These materials are promising to study slightly-doped multi-orbital Mott insulator.\cite{pandey_bamn2as2doped_2012}  
However, the evidence for strong-correlation physics in the FeSC does
can not rely on the presence of actual Mott states directly connected with
the superconducting compounds. Indeed no Mott insulator exists in the 122 family (BaFe$_2$As$_2$ or isoelectronic compounds doped
either with holes and electrons), the 1111 family which originates
doping LaFeAsO, the 11 selenides FeSe and FeTe, and LiFeAS. While this experimental fact is indeed completely
compatible with the scenario based on the Hund's coupling, it may cast
doubts on the whole relevance of correlations. 

%There have been efforts both in experimental as well as theoretical
%works to search for Fe-based compounds with electronic properties
%catching up above mentioned Mott physics, in order to bridge the gap
%with copper-based superconducting compounds. 
%
%A front of the research has been addressed to the search of Fe-based compounds with electronic properties catching up Mott physics to make a bridge with copper-based superconducting compounds.
%with manganese similarly to
%Fe-based materials have been synthesized in the search of novel
%antiferromagnetic Mott insulating states as in pristine cuprates.  

A close relative of FeSC with insulating behavior is
La$_2$O$_3$Fe$_2$Se$_2$. This system is based on a square
lattice of Fe ions with nominal valence 2+ as in all of iron
pnictides and chalcogenides. However, the resistivity as a function of
temperature shows clearly an insulating behavior with an activation 
energy gap of about 0.19\,eV.\cite{zhu_la2o3fe2se2_2010}   
On the basis of this experimental evidence combined with
Density-Functional Theory (DFT) supplemented by mean-field treatment
of the Hubbard U (DFT+U) in the
ordered magnetic phase it has been argued that La$_2$O$_3$Fe$_2$Se$_2$
is in a Mott state with low-temperature antiferromagnetic ordering.\cite{zhu_la2o3fe2se2_2010}
This evidence raises the natural question about the reason why the
Hund's metallic behavior of undoped FeSC is replaced with a Mott state
in this compound. 
A simple bandwidth reduction is a very unlikely answer, since one of
the defining properties of the Hund's metal regime is that the
critical value for a Mott transition is pushed to very large values,
significantly far from the experimental
estimates.\cite{georges2013_hund} This calls for the identification of
specific aspects of the bandstructure which make the present
material insulating.

In this work we investigate the differences of the electronic
structure of La$_2$O$_3$Fe$_2$Se$_2$ with respect to parent compounds
of the FeSC using the reference case of FeSe. We identify in a strong
lifting of the orbital degeneracy the main difference of the material
under consideration with respect to most parent compounds of
iron-based superconductors. This leads to a marked tendency towards
orbital-selective Mott transition and opens the way to a complete Mott
localization when combined with the reduced
kinetic energy of La$_2$O$_3$Fe$_2$Se$_2$ with respect to FeSe.

The paper is organized as follows. In Sec. II we present the
electronic structure which emerges from Density-Functional Theory
calculations, while In Sec. III we consider the inclusion of
electron-electron interactions. The section is organized in two
subsections dedicated respectively to the results obtained within the
slave-boson mean-field scheme and the more accurate Dynamical
Mean-Field Theory. Sec. IV contains our conclusions.

\section{Density-Functional Theory Bandstructure}

\begin{figure}
\includegraphics[width=.875\columnwidth,angle=-0]{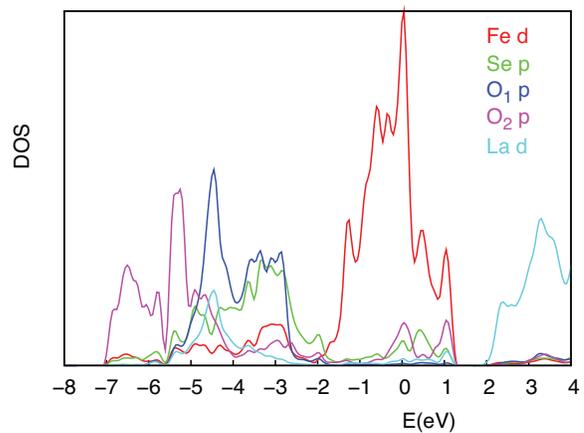}
\vspace{0.1cm}
\caption{(Color online) Density of states of
  La$_2$O$_3$Fe$_2$Se$_2$ calculated within PBE and 
  projected on the different atomic species. The zero energy is set at the Fermi level. } 
\label{fig1}
\end{figure}

We start our investigation determining the bandstructure of
La$_2$O$_3$Fe$_2$Se$_2$ by means of DFT in the framework of the generalized gradient approximation (using the
Perdew-Burke-Ernzerhof (PBE) functional\cite{perdew_generalized_1996}) for the tetragonal 
unit cell of La$_2$O$_3$Fe$_2$Se$_2$ \cite{zhu_la2o3fe2se2_2010} using
Quantum Espresso\cite{QE} and Wien2K\cite{blaha_wien2k_2001}. 
In Fig. \ref{fig1} we show the density of states projected on the
different atoms. 

PBE calculations clearly lead to a metallic solution with a sizable
spectral weight at the Fermi level. This low-energy  contribution to
the spectral density is dominated
by the bands arising from  Fe 3d electrons, which are very weakly
entangled with oxygen and lanthanum bands lying in energy windows
far from the Fermi level. The 3d  bands have an overall
width of 3.2 eV, considerably narrower than the 4.6 eV of the same
bands in FeSe\cite{aichhorn_theoretical_2010,lanata_orbital_2013}.  

In order to include the on-site Coulomb interaction parameterized by
the Hubbard $U$ and the Hund's coupling $J_h$ we
compute maximally localized 
Wannier orbitals\cite{wannier90}  for the pure 3d Fe orbitals built
from the iron bands in the energy range between -2  and 1.2\,eV. The
properties of these orbitals will also provide us with important information
that will help us to rationalize the behavior of  this materials.

The on-site energies of the Wannier orbitals reflect indeed an important difference between the
material under consideration and the parent compounds of the iron-based superconductors.
In La$_2$O$_3$Fe$_2$Se$_2$, the Fe ions are surrounded by two nearby
oxygen anions and four distant selenium anions.  
The additional oxygen ions result in a lower symmetry in
La$_2$O$_3$Fe$_2$Se$_2$ as compared to FeSe and other FeSC. This means that the
local problem is no longer diagonal in the standard cubic basis defined by the $e_g$ and $t_{2g}$ orbitals.
We can obviously diagonalize the local Hamiltonian for 
La$_2$O$_3$Fe$_2$Se$_2$ by means of a unitary transformation. 
The resulting orbitals, which are
linear combinations of the $(3z2-r2, xz, yz, x2 - y2, xy)$
orbitals\cite{wang_la2co2Se2O3_2010} that we label as $(1,2,3,4,5)$ have in
our calculations on-site energy of
$(-0.660,-0.627,-0.397,0.183,0.543)$\,eV, respectively.
This leads to a total crystal-field splitting of 1.1\,eV, 
much larger than the value for FeSe, which we estimate in 0.48\,eV
within analogous PBE calculations.

\begin{figure}
\includegraphics[width=.875\columnwidth,angle=0]{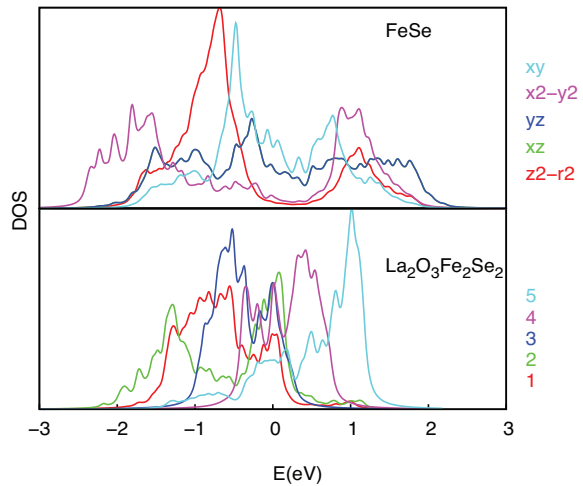}
%\vspace{0.2cm}
\caption{(Color online) Orbital-resolved density of states for
  La$_2$O$_3$Fe$_2$Se$_2$ (bottom panel) and FeSe (top panel). The zero energy is set at the Fermi level. } 
\label{fig2}
\end{figure}

Comparing the orbital-resolved density of states for
La$_2$O$_3$Fe$_2$Se$_2$ and FeSe (see Fig. \ref{fig2}) we can
visualize the different crystal-field splitting and emphasize another related effect: 
in FeSe the different orbitals have a similar width and lie more or less
in the same energy range. In contrast La$_2$O$_3$Fe$_2$Se$_2$ features
a set of narrow orbitals significantly shifted in energy relative to each
other. That means that the effective band-width reduction for each individual orbital
is even larger than the factor estimated from the total band width.
Interestingly, even in FeSe and other standard FeSC it
has been shown\cite{miyake_comparison_2010} that, arbitrarily
neglecting interorbital hybridizations turns the broad orbitals into
more localized objects, similar to what we find for
La$_2$O$_3$Fe$_2$Se$_2$. Therefore we can link the peculiar
orbital-resolved density of states of the latter material with the
crystal-field splitting that reduces the effect of the interorbital
hybridization. 

In the following of the paper we investigate how the difference in the
single-particle spectra reflects on the effects of electron-electron
correlations, keeping in mind the role of the Hund's coupling that we
anticipated above. The results picture can be quite rich because of
the non-trivial interplay between the ``Hund's physics'' and the two
main differences we highlighted, a reduced kinetic energy and an
increased crystal-field splitting. In particular, while a reduced
kinetic energy simply leads to effectively larger Coulomb terms, the
crystal-field splitting competes with the Hund's coupling, as the
latter tries to spread the electrons among the different orbitals to
maximize the total spin, while a large crystal-field splitting
obviously favors an unbalanced population with large occupation of the
low-lying orbitals.

\section{Effect of Electron-Electron Correlations}

To understand the role of electron-electron interactions in turning 
La$_2$O$_3$Fe$_2$Se$_2$ insulating we consider two different
approaches to treat the short-range interactions. We start from a
Slave-Spins mean-field (SSMF) theory \cite{demedici_slavespins_2005},
which allows for a computationally inexpensive and fast survey of the phase
diagram and it is expected to capture the main physics as long as the
system remains in a Fermi-liquid state\cite{demedici_selective2014}.
Then we move to the more accurate DFT+Dynamical Mean-Field Theory
(DMFT)\cite{georges_dynamical_1996} method, which treats exactly the
local quantum dynamics mapping the lattice model onto an impurity
embedded in a self-consistent bath. As impurity solver we employ
mainly Exact Diagonalization (ED).\cite{caffarel,capone_edsolver_2007}
We verified  for selected parameter the excellent agreement of ED with the continuous-time
Quantum Monte Carlo solution of the impurity model implemented in the TRIQS
toolkit\cite{ferrero_triqs,aichhorn2009,aichhorn2011}.

For the interacting part of the Hamiltonian we use a Kanamori form
parameterized by $U$ and $J_h$ according to

\begin{eqnarray}
H&_{int} = &  U\sum_{i,m}n_{im{\sigma}} n_{im{\sigma}'} +U'\sum_{i,m,m'}n_{im{\sigma}} n_{im'{\sigma}'} +\nonumber\\
&+&U^{''}\sum_{i,m,m'}n_{im{\sigma}} n_{im'{\sigma}} +\nonumber\\
& -&J_h \sum_{i,m,m'} [
d^{+}_{im{\uparrow}}d^{+}_{im'{\downarrow}}d_{im{\downarrow}}d_{im'{\uparrow}}
+\nonumber\\
& +
&d^{+}_{im{\uparrow}}d^{+}_{im{\downarrow}}d_{im'{\uparrow}}d_{im'{\downarrow}}
]
\label{Hamiltonian}
\end{eqnarray}

where $d_{i,m{\sigma}}$ is the destruction operator of an electron of
spin $\sigma$ at site i in orbital m, and $n_{im{\sigma}}=
d^{+}_{im{\sigma}}d_{im{\sigma}}$, $U$ and $U'=U-2J_h$,
$U^{''}=U-3J_h$ are intra- and inter-orbital repulsions and $J_h$ is
the Hund's coupling. 
In the absence of estimates of $U$ and $J_h$ for this material, we use the
constrained random-phase-approximation (cRPA) for 
FeSe ($U=4.2$\,eV and $J_h=0.504$\,eV) which are not expected to
differ in a critical way.\cite{miyake_comparison_2010}
Note that these interaction values are given in Hubbard-Kanamori
notation. The use of the same parameters also allows us to highlight
the role of the material-specific properties (kinetic energy and
crystal-field splitting) in determining the low-energy properties of
the system. In all the calculations we consider paramagnetic solution
that preserve also the orbital symmetry.

\subsection{Slave-spin Mean-Field Approximation}
In Fig. \ref{fig3} we show the DFT+SSMF orbital resolved densities $n_m$ and quasiparticle
weights $Z_m$  as functions of Coulomb parameter $U$ at fixed ratio
$J_h/U$= 0.2. Within this approximation spin-flip and pair-hopping
terms (last two lines of Eq. (\ref{Hamiltonian})) are not included. $Z_m$ is the measure of
the metallic character of each orbital. A vanishing of $Z_m$ for a given
orbital implies that the carriers with that character are reaching a
Mott localization. A Mott insulating state is reached when all the
$Z_m$'s vanish, while a situation with coexisting finite and zero $Z_n$'s
would correspond to an orbital-selective Mott transition (OSMT).

For small interactions, up to $U \simeq 1.5 eV$, the orbital densities range from 0.22 to 0.8
%%%%%%%(please put real numbers)
 and the $Z_m$ are close to one. Increasing
$U$, the $Z_m$ decrease and depart from each other while the orbital populations
deviate substantially from the non-interacting values. 
More precisely, the orbitals labeled as 4, 3 and 5 move, one after
the other, towards a
half-filled configuration ($n_m=0.5$) and they become insulating
($Z_m=0$) at different interaction strengths. In other words, the system shows a series of OSMT's
in which the different orbitals become localized independently on the
behavior of the others.
The two remaining orbitals (1 and 2)  instead remain metallic for a larger range
of interactions and they simultaneously become insulating for $U_c =
3.6 eV$, smaller than the estimated value for FeSe. At this critical interaction also orbital 2 becomes half-filled, while orbital 1 is
completely filled.
The ``order'' in which the different orbitals undergo a Mott
transition is a consequence of the individual bandwidth, which is
smaller for orbitals 4 and 3 and of the initial orbital population.

The series of transitions that we described is in stark contrast with
the results for FeSe\cite{lanata_orbital_2013} and for the other
parent compounds\cite{YuSi_LDA-SlaveSpins_LaFeAsO, demedici_selective2014}, where no OSMT occurs
despite a marked differentiation between the $Z_m$ of the different orbitals. In this case a full Mott
transition in which all the orbitals become simultaneously localized
takes place for values of the interactions which are much larger than
reasonable estimates for this material. In particular, for the cRPA values (for FeSe) introduced above,
La$_2$O$_3$Fe$_2$Se$_2$ is a Mott insulator, while FeSe is a Hund's
metal with small and orbital-differentiated quasiparticle weights.

It is useful to notice  that, increasing the value of $J_h$ the critical
coupling for an OSMT is reduced, while the critical U for a full Mott
transition increases \cite{demedici_lifting_2009}. This result,
together with the whole picture we have drawn, is in perfect agreement
with model calculations in which orbitals with the same bandwidth have been shifted leading to
an OSMT\cite{demedici_lifting_2009,demedici_genesis}.

Interestingly, while the full Mott localization requires a
commensurate filling, the OSMT that we observed survive also doping
the $d^6$ Mott insulator. In the inset of Fig. \ref{fig3} we show the
evolution of the quasiparticle weight as a function of $U$ for a
$d^{5.5}$ configuration, where we doped one hole every two iron
sites. 
Comparing with the $d^6$ case, we see that orbitals 3, 4 and 5 still
undergo the series of OSMT's we described above, and in particular the
three orbitals are insulating for the $U$ values representative of
La$_2$O$_3$Fe$_2$Se$_2$, while orbitals 1 and 2 become metallic after doping.
This can be explained in terms of a schematic general picture of OSMT
\cite{demedici_book}. 
For sizable $J/U$ the global half-filled configuration is in general a Mott insulator at quite smaller U than for the $d^6$ case. 
In this $d^5$ Mott insulating phase each orbital opens an independent
gap, thus being half-filled when the chemical potential falls within
its gap.
The gaps can have different widths and position depending on the
crystal-field splitting and the bandwidth of their orbital. Doping of
a Mott insulator 
occurs essentially when the chemical potential moves out of the gap, and thus
in this situation this happens for each orbital at a separate value of
the chemical potential. The selective doping of one or more orbitals
leads therefore to an OSMT.
Our results show that the bandstructure of  La$_2$O$_3$Fe$_2$Se$_2$ is
such that when reaching a total filling of 6 electrons only two
orbitals  (1,2) have actually been doped compared to the half-filled case while the rest remain half-filled, have an open gap, and are thus insulating. This does not depend on the filling being exactly 6 or slightly above or below, thus doping the $d^6$ configuration tunes indeed the Mott transition for orbitals 1 and 2 but does not alter the selective insulating behavior for orbitals 3,4,5. 

Our results strongly indicate that the main anomaly of
La$_2$O$_3$Fe$_2$Se$_2$ with respect to the parent compounds of FeSC is the strong enhancement of the
crystal-field splitting with respect to iron-based superconductors,
which conjures with the Hund's coupling to decouple the different
orbitals, leading to an orbital selective localization rather than to
a global Mott transition for the six electrons. The overall reduction
of the kinetic energy with respect to FeSe plays a more quantitative
role, reducing the actual critical value of  $U$. 

\subsection{Dynamical Mean-Field Theory}
Our DFT+SSMF treatment is however limited to metallic solutions and it
can not help us to characterize the Mott insulating solution for $U >
U_c$. A full description of both low- and high-energy features can
instead be obtained by means of DFT+DMFT. As mentioned above, we
mainly use and ED solution for the impurity model, which also allows
us to include pair-hopping and spin-flip terms, as opposed to DFT+SSMF.

\begin{figure}
\includegraphics[width=.99\columnwidth,angle=-90]{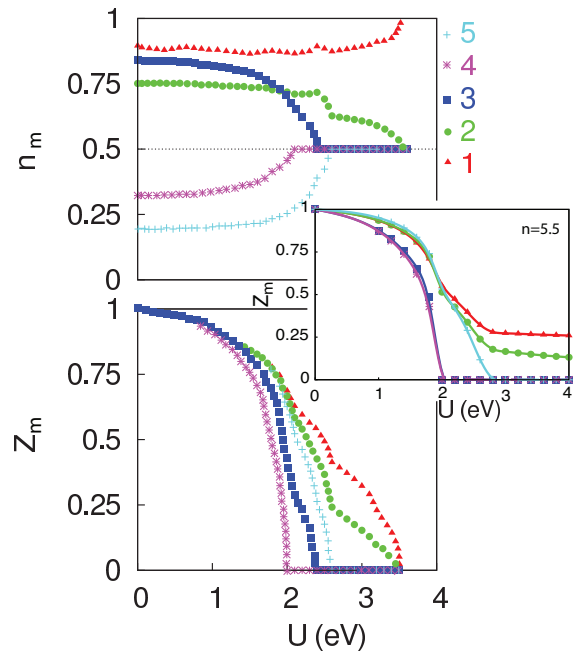}
\caption{(Color online) Orbitally resolved density ($n_m$) and
  quasiparticle weights ($Z_m$) calculated within DFT + SSMF for
  La$_2$O$_3$Fe$_2$Se$_2$ at ratio $J_h/U=0.20$. The inset shows $Z_m$
  for a filling of $d^{5.5}$ electrons per iron (one hole every two
  iron sites).} 
\label{fig3}
\end{figure}

We consider  $N_s=15$ orbitals in total, with
5 impurity orbitals and 2 bath degrees of freedom connected to each
impurity orbital.  
The storage requirements for the matrix hamiltonian and Lanczos
vector, since the Hilbert space has dimension 41 409 225, are solved by
splitting them over parallel processors. The impurity solver is
diagonalized by a parallel Arnoldi algorithm \cite{ARPACK} using the
symmetry with respect to the inversion of up and down particles. 

In Fig. \ref{fig3} we show the dynamical information obtained for the
cRPA values $U=4.2$\,eV and $J_h=0.504$\,eV. We present orbital-resolved Green's functions 
on the imaginary-frequency axis (a)), which is the most direct
product of the calculation, and the real-frequency spectral function (b)),
which provides a more obvious physical content.
The imaginary part of the Matsubara Green's functions smoothly
extrapolate to zero in the limit of small frequency. This clearly
confirms that the material is insulating also within the more accurate
DFT+DMFT method. The imaginary part of the self-energy diverges only for orbital 5, for which the chemical potential
happens to lie in the middle of the gap (See also the spectral
function in panel b)). For the other orbitals the chemical potential
lies far from the center of the gap and, consequently, the imaginary
part of the self-energy does not diverge. Finally, a comparison of the
orbital-projected spectral density with the DFT results of Fig. \ref{fig1}
allows us to visualize the
formation of high-energy spectral weight due to strong correlations
and to estimate the  Mott gap $\Delta_{Mott} \simeq 1 eV$. The Mott
insulating state of La$_2$O$_3$Fe$_2$Se$_2$  is characterized by a
high-spin configuration with $S=2$ as a consequence of the Hund's
coupling.
The orbital populations are in agreement with the DFT+SSMF results,
confirming the picture of four half-filled orbitals and one full
orbital. Therefore the whole scenario obtained within DFT+SSMF survives
when we move to the full DMFT treatment of correlations, suggesting
that also the sequence of OSMT transition preceding the full Mott
localization is a real feature of the present material. 
The disagreement between the theoretical value of the gap and
experiments may be attributed to an overestimate of $U$ and $J_h$, or
even in their ratio. However, as mentioned above, the choice to use
the same values of FeSe allowed us to highlight the difference between
La$_2$O$_3$Fe$_2$Se$_2$ and the parent compounds of standard FeSC.

All the calculations we presented are limited to a paramagnetic
solution without broken symmetry in any channel, demonstrating the
intrinsic Mott character of the insulating state. However, at low
temperature this state may be unstable towards different ordering as
magnetic and orbitals ones\cite{giovannetti_bacro3_2014u,greger_2013u}
as it is indeed suggested by DFT+U calculations \cite{zhu_la2o3fe2se2_2010}.

To confirm the reliability of our results, we compared with a
different DFT+DMFT implementation based on the TRIQS
package.\cite{ferrero_triqs,aichhorn2009,aichhorn2011} In these
calculation we use
projective Wannier function techniques, as well as continuous-time
quantum Monte Carlo as impurity solver. In addition, we considered also the
Se p states for the construction of the Wannier functions, in order to
check the stability of the insulating state. For the  interactions we use
also cRPA values for FeSe, and a Slater expansion of the interaction
hamiltonian \cite{aichhorn_theoretical_2010}.

The results confirms the precedent picture. The system is in
a Mott state, with four orbitals close to half-filling, and one completely
filled. The insulating state appears therefore extremely solid, as it does
not depend on the details of the computational scheme that is applied.

\begin{figure}
\includegraphics[width=.625\columnwidth,angle=-90]{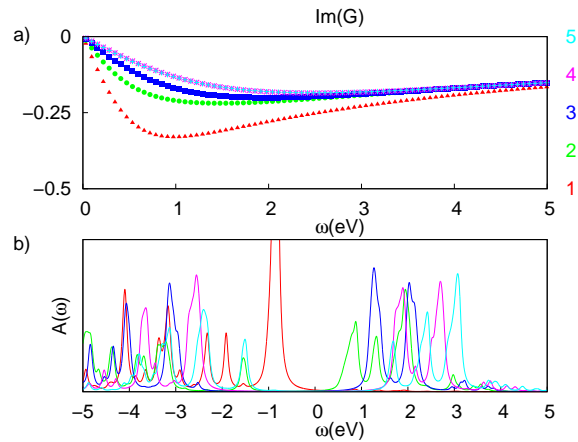}
\vspace{0.5cm}
\caption{(Color online) a) Orbital-resolved imaginary part of local
  Green's function (G) as a function of the Matsubara imaginary frequency,
b) spectral density (A) as a function of the real frequency computed within DFT + DMFT for La$_2$O$_3$Fe$_2$Se$_2$.} 
\label{fig4}
\end{figure}

\section{Conclusions}
In conclusion, we use different approaches combining DFT with strong
correlation physics to study the electronic
properties of the new synthesized compound
La$_2$O$_3$Fe$_2$Se$_2$, an insulating material that shares the same electron
count with the metallic parent compounds of iron superconductors. Our
calculations clearly demonstrate that the material is a Mott insulator.
A detailed understanding of the electronic structure of La$_2$O$_3$Fe$_2$Se$_2$ compared to the one of FeSe 
highlights the difference of this class of compounds with ordinary
Fe-based superconducting materials. Using the same interaction values
for the two materials, we demonstrate that the insulating behavior of
La$_2$O$_3$Fe$_2$Se$_2$ is not simply due to a larger interaction
strength, but it is related to a more fundamental difference in the
electronic structure.

We identify the main reason for the insulating behavior in the
enhancement of the crystal-field splitting due to the position of the
oxygen atoms. The larger crystal-field leads to a reduced overlap
between the density of states with different orbital character, which
triggers a series of successive Mott transitions in which the
different orbitals become insulating one after the other as the
interaction grows. The insulator is indeed characterized by four
singly occupied orbitals and a fully occupied one. This condition is
quite different from iron-based superconductors, in which a marked
orbital differentiation does not however lead to individual Mott
insulating behavior.

As a consequence, we predict that La$_2$O$_3$Fe$_2$Se$_2$ should give
rise to a series of OSMT under pressure and we expect that a similar
behavior should take place as a function of doping. A slightly doped
material is expected to show the coexistence between localized
orbitals and delocalized orbitals hosting the extra charges. 

Note Added. During completion of this work, we became aware of a related work by B. Freelon et Al., 
which is complementary to ours and gives comparable results.

\section*{Acknowledgments}
GG and MC are financed by European Research Council
under FP7/ERC Starting Independent Research Grant ``SUPERBAD" (Grant
Agreement n. 240524). 
%MA is supported from the SFB VICOM sub-project F04103. 
MA is supported from the Austrian Science Fund (FWF) through projects F04103, P26220, and Y746.
Calculations have been perfomed at  CINECA within
the HPCproject 2014 CONDMAG (lsB04) and the TU Graz
dcluster.

\bibliography{Iron_Superconductors}

\end{document}